\documentclass[aps,epsfig,twocolumn]{revtex4}
\usepackage{amsmath}
\usepackage{epsfig}
\usepackage{graphics}
\begin{document}
\title{Bragg Spectroscopy of Cold Atomic Fermi Gases}
\author{G.\ M.\ Bruun$^1$ and Gordon\ Baym$^{2,3}$}
\affiliation{$^1$Niels Bohr Institute, Blegdamsvej 17, 2100 Copenhagen,
Denmark\\
$^2$NORDITA, Blegdamsvej 17, 2100 Copenhagen, Denmark\\
$^3$Department of Physics, University of Illinois, 1110 W.\ Green St.,
Urbana, IL 61801}

\date{\today{}}

\begin{abstract}

    We propose a Bragg spectroscopy experiment to measure the onset of
superfluid pairing in ultracold trapped Fermi gases.  In particular, we study
two component Fermi gases in the weak coupling BCS and BEC limits as well as
in the strong coupling unitarity limit.  The low temperature Bragg spectrum
exhibits a gap directly related to the pair-breaking energy.  Furthermore, the
Bragg spectrum has a large maximum just below the critical temperature when
the gas is superfluid in the BCS limit.  In the unitarity regime, we show how
the pseudogap in the normal phase leads to a significant suppression of the
low frequency Bragg spectrum.

\end{abstract}

\maketitle

%-Checked f-sum rule. <0.01 error for normal state. <0.03 Obeys!

\section{Introduction}

    Bragg spectroscopy, using two laser beams, has proven to be a very
successful tool for probing the structure factor of trapped atomic
BEC's~\cite{BraggExp}.  In this paper, we examine the analogous experiment on
two-component Fermi gases in both the normal and superfluid phases.  We
calculate the Bragg scattering rate for varying interaction strengths in both
the weak coupling BCS and BEC limits and in the strong coupling unitarity
limit.  Previously, we proposed measuring the inelastic (Stokes/anti-Stokes)
scattering of an off-resonant laser beam on a two-component Fermi
gas~\cite{BruunBaym}.  In the BCS limit, the intensity of the scattering in
this Stokes experiment exhibits a large maximum just below the critical
temperature, $T_c$, at which the gas becomes superfluid.  This effect is the
light scattering analog of the Hebel-Slichter effect in conventional
superconductors, a hallmark experimental test of BCS theory~\cite{Hebel}.
Here, we demonstrate that the same effect can be observed in a Bragg
scattering experiment.  We furthermore show that the pairing induced by the
interactions can be readily detected in the Bragg spectrum.  In the strong
coupling regime, we discuss how the presence of a pseudogap in the normal
phase has significant effects on the Bragg spectrum.  The advantage of the
Bragg experiment as compared to the Stokes experiment is that one varies in a
controlled way the energy and momentum imparted to the atoms.  Compared with
recent radio frequency (rf) experiments which involve a third unoccupied
hyperfine state~\cite{Chin}, the Bragg experiment involves only the two
hyperfine states already present in the trap.  The Bragg experiment, avoiding
complications due to non-trivial interaction effects between the third
hyperfine state present in the rf experiment \cite{Yu} and the trapped gas, is
thus simpler to interpret.

\section{Bragg Scattering}

    We consider a homogeneous gas of fermionic atoms occupying two hyperfine
states, denoted by $|\uparrow\rangle$ and $|\downarrow\rangle$.  We assume
that the state $|\downarrow\rangle$ has energy $\omega_{\rm hf}$ ($\hbar=1$
and $k_B=1$ in this paper) above the state $|\uparrow\rangle$, and that the
population of the two states is equal, $N_\uparrow=N_\downarrow=N/2$.  The two
atomic states interact with a coupling which can be characterized by a scattering
length $a$ in a vacuum. We ignore the interaction between atoms in the same
hyperfine state since $s$-wave scattering is suppressed between two identical fermions. 

\begin{figure}
\includegraphics[width=0.6\columnwidth,height=0.4\columnwidth,
 angle=0,clip=]{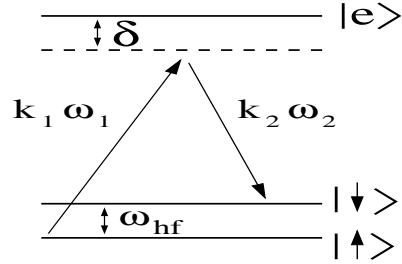}
\caption{Schematic atomic level diagram for stimulated Bragg scattering
  leading to a ``spin" flip.}
\label{BraggFig}
\end{figure}

    Bragg scattering is realized by a pair of laser beams with wave vectors
${\mathbf{k}}_1$ and ${\mathbf{k}}_2$ and frequencies $\omega_1$ and
$\omega_2$ illuminating the gas.  An atom absorbing a photon from beam 1 and
emitting a photon in beam 2 changes its energy-momentum by
$\omega_1-\omega_2,{\mathbf{k}}_1-{\mathbf{k}}_2$.  Bragg scattering can be
used to probe both the density and spin correlation functions.  Pairing is
predicted to have several effects on the frequency and momentum dependence of
these correlation functions for $T=0$~\cite{Buchler,Deb,Mihaila}.  To probe
the spin-flip correlation function, one tunes the two laser frequencies so
that $\omega_1-\omega_2\sim\omega_{\rm hf}$ as illustrated in
Fig.~\ref{BraggFig}.

    If the detuning, $\delta$, from the optical transition to or from the
upper level $|e\rangle$ is large, population of the upper level can be
neglected, and the effective light matter Hamiltonian describing spin-flip
Bragg scattering can be written as
\begin{equation}
   H_{\rm B}=\int d^3 r\left[{\mathcal{I}}
 ({\mathbf{r}},t)\psi_\downarrow^\dagger({\mathbf{r}},t)
 \psi_\uparrow({\mathbf{r}},t)+\rm{h.c.}
 \right],
 \label{BraggTerm}
\end{equation}
where ${\mathcal{I}}({\mathbf{r}},t)=e^{-i(\omega_1-\omega_2)
t}{\mathcal{I}}({\mathbf{r}})$, and ${\mathcal{I}}({\mathbf{r}})$ describes
the spatial profile of the laser beams and the dependence on the atomic states
involved and their coupling to the electromagnetic field~\cite{Carusotto}.  We
neglect finite waist effects and take
${\mathcal{I}}({\mathbf{r}})={\mathcal{I}}e^{i{\mathbf{q\cdot r}}}$ with
${\mathbf{q}}={\mathbf{k}}_1-{\mathbf{k}}_2$.

    To measure the effects of pairing we focus on the spin-flip rate
$\langle{\dot{N}}_\downarrow\rangle$ induced by the Bragg lasers, where
$N_\downarrow$ is here the number of particles in the level
$|\downarrow\rangle$, and $\langle \ldots\rangle$ denotes a thermal average.
The Heisenberg equation of motion for the operator $N_\downarrow$ is
\begin{equation}
  i\dot{N}_\downarrow=[N_\downarrow,H_{\rm B}]=
  \int d^3 r\left[{\mathcal{I}}({\mathbf{r}},t)
  \psi_\downarrow^\dagger({\mathbf{r}},t)
  \psi_\uparrow({\mathbf{r}},t)-\rm{h.c.}\right].
\end{equation}
For sufficiently weak laser beams, $\langle\dot{N}_\downarrow\rangle$
can by calculated by linear response theory:
\begin{gather}
 \langle\dot{N}_\downarrow\rangle=
 i{\mathcal{I}}^2\int\frac{d\omega'}{2\pi}
 \left(\frac{\langle[\psi_\uparrow^\dagger
 \psi_\downarrow,\psi_\downarrow^\dagger
 \psi_\uparrow]\rangle({\mathbf{q}},\omega)}
 {\omega-\omega'}-c.c.\right)\nonumber\\
 =-2{\mathcal{I}}^2{\rm Im}\mathcal{D}({\mathbf{q}},\omega+i\eta),
\label{Braggterm}
\end{gather}
with $\omega=\omega_1-\omega_2-\omega_{\rm hf}$.  Bragg scattering thus
probes the Fourier transform of the spin-flip correlation function
$\langle[\psi_\uparrow^\dagger({\mathbf{r}},t)
\psi_\downarrow({\mathbf{r}},t),\psi_\downarrow^\dagger ({\mathbf{r}}',t')
\psi_\uparrow({\mathbf{r}}',t')]\rangle$.  The response of the gas is usually
detected experimentally by measuring the total momentum imparted on the
gas~\cite{BraggExp}; the rate of momentum transfer is straightforwardly
related to the scattering rate by
\begin{equation}
 \dot{\mathbf{P}}={\mathbf{q}}\dot{N}_\downarrow
\end{equation}
where ${\mathbf{P}}={\mathbf{P}}_\downarrow+{\mathbf{P}}_\uparrow$ is the
total momentum of the gas in the two hyperfine states.

\section{Weak coupling BCS limit}

    In the weak coupling limit $a\to 0^-$, the gas can be described by BCS
theory.  The correlation function in (\ref{Braggterm}) is then a sum of
contributions from normal and anomalous Green's functions~\cite{Fetter}.  We obtain after
some algebra,
\begin{gather}
 \mathcal{D}_0(q,z)=\int\frac{d^3k}{(2\pi)^3}
 \left[(uu'+vv')^2\frac{f-f'}{z+E-E'}+\right.\nonumber\\
 \left.\frac{1}{2}(uv'-vu')^2\left(\frac{(1-f-f')}{z-E-E'}
 +\frac{(f+f'-1)}{z+E+E'}\right)\right],
 \label{uvexpression}
\end{gather}
with $z=\omega+i\eta$.  Here, $u=u_k$ and
$u'=u_{{\mathbf{k}}+{\mathbf{q}}}$, etc., with $u_k^2=(1+\xi_k/E_k)/2$ and
$v_k^2=1-u_k^2$ as usual in BCS theory.  The quasiparticle energies are
$E=E_k=\sqrt{\xi_k^2+\Delta^2}$ and $E'=E_{\bf k+q}$ with
$\xi_k=\epsilon_k-\mu$, where $\epsilon_k=k^2/2m$, $\mu$ is the chemical
potential, and $\Delta$ the BCS pairing gap.  The Fermi functions are $f=f(E)$
and $f'=f(E')$, with $f(x)=\left(e^{x/T}+1\right)^{-1}$.  The first term of
(\ref{uvexpression}), describes scattering of quasiparticles while the last
two terms describe the creation and annihilation of two quasiparticles -- pair
breaking processes.

    For an ideal gas in the normal phase, (\ref{uvexpression}) yields in the
limit $\omega\ll\epsilon_F$, $T\ll T_F$ ($T_F$ is the Fermi temperature), and
$\omega\ll qk_F/m$,
\begin{equation}
 {\rm{Im}}\mathcal{D}_0(q,\omega)=-\frac{3\pi}{16}
 \frac{nk_F\omega}{q\epsilon_F^2},
 \label{Nexact}
\end{equation}
where $\epsilon_F$ is the Fermi energy, $n=k_F^3/3\pi^2$ is the total
density of the gas, and $k_F$ is the Fermi momentum.

    The spin flip induced by Bragg scattering can excite collective spin
waves.  These modes are undamped in the normal phase for an attractive
interaction~\cite{NegeleOrland}.  For a repulsive interaction in the normal
phase, the spin waves are damped and they change the spectral shape of the
particle-hole continuum.  The single particle approximation
(\ref{uvexpression}) does not include such collective modes.  In order to
include them, we sum bubbles to obtain the random phase (RPA) approximation~\cite{Fetter},
\begin{equation}
 \mathcal{D}_{\rm RPA}=
 \frac{\mathcal{D}_0}{1+\Gamma\mathcal{D}_0},
 \label{RPA}
\end{equation}
with $\Gamma=4\pi a/m$ the scattering matrix for $k_Fa\ll 1$.  Note the
$+$ sign in the denominator due to the spin flip.

\subsection{Results}

    We first consider the temperature dependence of the scattering rate.  An
attractive feature of the Stokes experiment is that one can observe a
strikingly large maximum in the scattered signal just below the critical
temperature $T_c$~\cite{BruunBaym}.  We now demonstrate that the same effect
is present for the Bragg scattering experiment in the experimentally
realizable parameter range, $\omega \ll \Delta$, $\omega < qk_F/m$, and $q<
2k_F$.

\begin{figure}
\includegraphics[width=0.8\columnwidth,height=0.6
  \columnwidth,angle=0,clip=]{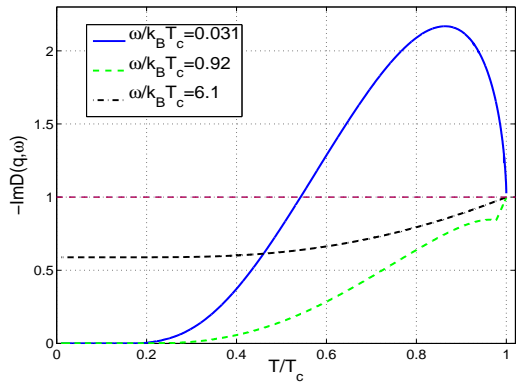}
\caption{The scattering rate $-{\rm{Im}}D(q,\omega)$ in units of the
scattering rate in the normal phase vs.  $T$ for representative $\omega$.
(color online).}
\label{TDepFig}
\end{figure}

    Figure\ \ref{TDepFig} shows the scattering rate (\ref{Braggterm}) as a
function of $T$ in the weak coupling limit, in units of the normal phase rate,
which is constant for $T\ll T_F$.  For illustration, we have taken $k_Fa=-0.3$
and $q=k_F$.  Solving the BCS equations as a function of $T$, we obtain the
scattering rate from Eqs.~(\ref{uvexpression}) and (\ref{RPA}).  The critical
temperature is $T_c=0.0033\epsilon_F$.  We plot the results for frequency
differences between the two laser beams $\omega=0.031T_c$, $0.92T_c$, and
$6.1T_c$.  The numerical normal phase results agree with (\ref{Nexact}),
confirming the numerical accuracy of the calculations.

    As Fig.\ \ref{TDepFig} shows, the scattering rate in the superfluid state
has a large maximum below $T_c$ for $\omega\ll T_c$.  This maximum is due to a
large quasiparticle scattering rate arising from the increased density of
states at the Fermi energy in the superfluid phase just below $T_c$, $N(E) =
(p_F/2\pi^2)E_k/\xi_k$. For $\omega\ll T_c$, both scattering states $\bf k$ and
$\bf{k+q}$ in Eq.~(\ref{uvexpression}) can be in the region around the Fermi
surface of a large density of states yielding an increased scattering rate.
The maximum in the scattering rate depends crucially on the coherence factors
for quasiparticle scattering in (\ref{uvexpression}) adding, viz.,
$uu'+vv'\sim 1$, for a spin flip.  For Bragg scattering into the same
hyperfine state, the coherence factor is $uu'-vv'\sim 0$ at the Fermi surface
and the increased density of states would not lead to a maximum in the
scattering rate.  For $T\to 0$, the scattering rate is suppressed by the
increasing gap $\Delta(T)$ since the density of quasiparticles available for
scattering scales as $\exp(-\Delta/T)$.  For $\omega<2\Delta$ the lasers
cannot break Cooper pairs. 

    The physical reason for the coherence factors adding ($uu'+vv'\sim 1$) for
spin-flip scattering is that the quasiparticles carry the same spin in the
superfluid as in the normal phase.  On the other hand, in scattering without
spin flip, which couples to the particle density, the particle number carried
by the quasiparticles is suppressed at the Fermi surface, since a
quasiparticle is a superposition of a hole and a particle.  In this case the
coherence factors subtract ($uu'-vv'$).

    For increasing $\omega$, the peak in the rate below $T_c$ decreases and
eventually disappears for $\omega\sim T_c$.  From Fig.\ \ref{TDepFig}, we see
that the scattering rate in the superfluid state is always smaller than in the
normal state when $\omega\gtrsim T_c$ (green dashed and black dash-dotted lines). 
 This is because both scattering states cannot be in the
region around the Fermi surface where the single particle density of states is
increased.  Bragg scattering does not probe the increased density of states
for the superfluid state for higher frequency as effectively as for $\omega\ll
T_c$, and the resulting rate is decreased compared to the normal state.  Note
that the kink in scattering rate just below $T_c$ for $\omega=0.92T_c$, at
which point $2\Delta(T)=\omega$, is physical.  Below this temperature, the gap
is too large to allow the creation of two quasiparticles and the second term
in (\ref{uvexpression}) vanishes.

    Finally, Fig.\ \ref{TDepFig}, for $\omega=6.1T_c$ illustrates the case of
large energy transfer, $\omega>2\Delta(T=0)$.  Contrary to the case
$\omega<2\Delta(T=0)$, the rate is finite for $T=0$ as the energy transfer is
large enought to break pairs and the second term in (\ref{uvexpression}) is
finite.  For $\omega/T_c\to \infty$, the scattering rate approaches that of
the normal state, as expected.

    We conclude that there is a peak in the scattering rate for $T<T_c$ and
$\omega\ll T_c$.  Also, the peak disappears unless $\omega<qk_F/m$ and
$q<2k_F$; otherwise, both scattering states cannot be in the region of
increased density of states.

\begin{figure}
\includegraphics[width=0.8\columnwidth,
height=0.6\columnwidth,angle=0,clip=]{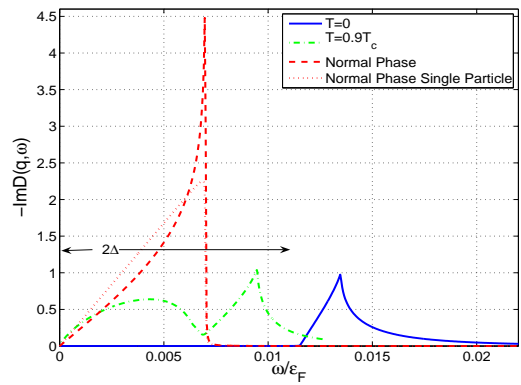}
\caption{The scattering rate $-{\rm{Im}}D(q,\omega)$ in units of
$n/2\epsilon_F$ as a function of $\omega$ for $k_Fa=-0.3$ and various
temperatures (color online).} \label{wDepFig}
\end{figure}

    We now consider the frequency dependence of the Bragg scattering rate.
Figure\ \ref{wDepFig} shows the rate from (\ref{RPA}) as a function of
$\omega$ for the gas in the weak coupling limit with $k_Fa=-0.3$ and
$q/k_F=0.0035$.  The scattering rate is depicted for various temperatures
below and just above $T_c$ where the gas is normal.  For comparison, we also
plot the result of the single particle approximation (\ref{uvexpression}) in
the normal phase.

    The large peak in the normal phase response just above the single particle
continuum arises from the spin wave present above the particle-hole continuum
for an attractive interaction.  The finite width of the peak is an artifact of
our having included a small imaginary part in the frequency $\omega$ for
computational reasons.  We see from Fig.\ \ref{wDepFig} that when the gas
becomes superfluid, the spin wave vanishes.  This is because only potential
flow, i.e., density fluctuation modes, is possible in the
superfluid~\cite{Leggett}.  Other modes such as spin waves require a normal
component to be present.  In a trapped system this effect leads to the damping
of the spin-dipole mode in the superfluid phase~\cite{BruunMottelson}.  In the
superfluid phase, the particle-hole spectrum disappears and is replaced by a
low frequency gap, which reflects the pair breaking energy described by the
second term in (\ref{uvexpression}).  At zero temperature the gas does not
respont for $\omega<2\Delta$, the minimum energy required to break a pair.

    The pairing gap can thus be detected directly as a gap of $2\Delta$ in the
Bragg spectrum.  By contrast, in the rf experiment of \cite{Chin} the shift in
the spectrum scales as $\Delta^2/\epsilon_F$, since one transfers particles to
a third hyperfine state which in general interacts differently with the two
paired hyperfine states~\cite{Yu}.

\section{Weak coupling BEC limit}

    We now consider Bragg scattering in the weak coupling BEC regime $k_Fa\to
0^+$.  In this limit, the molecules are tightly bound with a radius much
smaller than the average interparticle distance, and are weakly interacting
both with each other and with unbound atoms~\cite{Randeria}.  The system can therefore to a
good approximation be regarded as an ideal gas of point molecules and atoms in
thermal equilibrium.  The molecules Bose condense below the critical
temperature $T_c=0.218T_F$.  The Bragg scattering induced by (\ref{BraggTerm})
then either breaks a molecule by flipping the spin of a bound $\uparrow$ atom,
thereby creating two free $\downarrow$ atoms, or it simply flips the spin
of a free $\uparrow$ atom.  The Fermi Golden Rule result for Bragg scattering
rate into the free atomic $\downarrow$ states for such a system is then a sum
of three terms,
\begin{equation}
 \dot{n}_\downarrow={\mathcal{I}}^2(\Gamma_{\rm atom}+\Gamma_{\rm
 BEC}+\Gamma_{\rm Mol}).\label{MolBragg}
\end{equation}
The scattering from the unbound atoms is given by
\begin{equation}
 \Gamma_{\rm atom}
 =\int\frac{d^3k}{(2\pi)^3}(f-f')2\pi\delta(\omega+\epsilon-\epsilon'),
\label{AtomBragg}
\end{equation}
where $\epsilon=\epsilon_k$, $\epsilon'=\epsilon_{\bf k+q}$, $f=f(\xi)$,
and $f'=f(\xi')$.  The rate from a BEC of molecules with density $n_0$ is
\begin{equation}
 \Gamma_{\rm BEC}
 =n_0\int\frac{d^3k}{(2\pi)^3}|M|^2(1-f-f')
 2\pi\delta(\omega+E_b-\epsilon-\epsilon'),
 \label{BragBEC}
\end{equation}
with $E_b=-\hbar^2/ma^2$ the energy of a molecule with  zero momentum
relative to the bottom of the free particle continuum of a pair of $\uparrow$
and $\downarrow$ atoms, and $M=M({\bf k},{\bf q})$ the matrix element of
(\ref{BraggTerm}) (excluding ${\mathcal{I}}$) between a molecule at rest and a
pair of free $\downarrow$ atoms with momenta ${\mathbf{k+q}}$ and
$-{\mathbf{k}}$.  The scattering rate from thermal molecules is
\begin{gather}
 \Gamma_{\rm
 Mol}=\int\frac{d^3p}{(2\pi)^3}\int\frac{d^3k}{(2\pi)^3}|M|^2\nonumber\\
 [n_B(\epsilon_{\rm
 mol})(1-f-f')-ff']2\pi\delta(\omega+E_b+\frac{p^2}{4m}-\epsilon-\epsilon'),
 \label{BragTherm}
\end{gather}
with $n_B(x)=(e^{x/T}-1)^{-1}$ the Bose distribution, and $\epsilon_{\rm
mol}(p)=E_b+p^2/4m-2\mu$ the energy of molecules with momentum $p$.  The
molecule breaks into two $\downarrow$ atoms with momenta $\mathbf{k+q+p}/2$
and ${\mathbf{p}}/2-\mathbf{k}$ and energies
$\epsilon=\epsilon_{{\mathbf{p}}/2-\mathbf{k}}$ and
$\epsilon'=\epsilon_{\mathbf{k+q+p}/2}$.  The matrix element in
(\ref{BragTherm}) is, from Galilean invariance, independent of $\bf p$.

    To proceed, we model the molecule-free atom pair matrix element $M=M({\bf
k},{\bf q})$ as follows.  We write state of a molecule at rest as
$|\phi\rangle=\sum_{\mathbf{k}}\phi_kc^\dagger_{{\mathbf{k}}\uparrow}
c^\dagger_{-{\mathbf{k}}\downarrow}|0\rangle$, with $|0\rangle$ the vacuum and
$c^\dagger_{{\mathbf{k}}\sigma}$ creating an atom with momentum ${\mathbf{k}}$
in hyperfine state $\sigma$.  The Fourier transform of the molecular wave
function is $\phi_k$.  The spin-flip process illustrated in Fig.\
\ref{BraggFig} breaking the molecule and adding momentum ${\mathbf{q}}$
creating two free $\downarrow$ atoms with momenta ${\mathbf{k+q}}$ and
$-{\mathbf{k}}$ corresponds to $c^\dagger_{{\mathbf{k+q}}\downarrow}
c_{{\mathbf{k}}\uparrow}|\phi\rangle
=\phi_kc^\dagger_{{\mathbf{k+q}}\downarrow}
c^\dagger_{-{\mathbf{k}}\downarrow}|0\rangle$.  However, the process
$c^\dagger_{-{\mathbf{k}}\downarrow}c_{-{\mathbf{k-q}}\uparrow}|\phi\rangle
=-\phi_{|{\mathbf{k+q}}|}c^\dagger_{{\mathbf{k+q}}
\downarrow}c^\dagger_{-{\mathbf{k}}\downarrow}|0\rangle$ connects to the same
final state, and the two matrix element must be added coherently in the Golden
Rule expression.  We thus find the matrix element
$|M({\mathbf{k}},{\mathbf{q}})|^2=|\phi_k-\phi_{|{\mathbf{k+q}}|}|^2/2$ (where
the factor $1/2$ avoids double counting).

    The BCS wave function correctly describes a condensate of non-overlapping
bosons (Cooper pairs) with binding energy $E_b=-\hbar^2/ma^2$ in the BEC limit
$a\to 0^+$.  To connect this wave function to the above analysis, we note that
in the BCS expression the Bragg scattering (\ref{uvexpression}), it is the
second term that describes the breaking of Cooper pairs.  Thus, BCS theory
yields $|M({\bf k},{\bf q})|^2=(u_kv_{\bf k+q}-v_ku_{\bf k+q})^2/2$.  In the
BEC limit, we have $u_k\rightarrow 1$ and
$v_k\rightarrow\Delta/(2\epsilon_k+|E_b|)$ with $\Delta\rightarrow 2\sqrt{\pi
n}/(m\sqrt{a})$.  BCS theory therefore yields $\phi_k\rightarrow v_k$ for the
Fourier transform of the molecular wave function and
\begin{equation}
  \phi(r)=\sum_{\mathbf{k}}v_ke^{i{\mathbf{kr}}}\propto
  \frac{e^{-r/a}}{\sqrt{a}r}
\end{equation}
in the BEC limit.  This is the
asymptotic bound state wave function for a potential with scattering length
$a$ as expected.  We therefore use the BEC limit of the BCS result for the
matrix element $|M({\bf k},{\bf q})|^2$.  Note that we neglect any closed
channel components deep in the BEC regime.

    \subsection{Results}

\begin{figure}
\includegraphics[width=0.7\columnwidth,
height=0.5\columnwidth,angle=0,clip=]{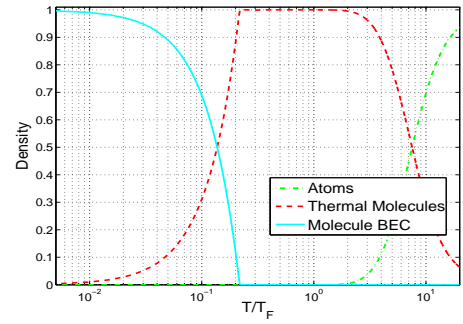}
\caption{Contributions to the total atom density (in units of $n$) from the
molecular BEC, the thermal molecules, and the free atoms.}
\label{DensFig}
\end{figure}

    The populations of the molecular and atomic states are found by solving
the atom-molecule equilibrium problem; the density of the thermal molecules,
the molecular BEC and the thermal atoms are shown in Fig.\ \ref{DensFig} for
varying $T$.

\begin{figure}
\includegraphics[width=0.7\columnwidth,
height=0.5\columnwidth,angle=0,clip=]{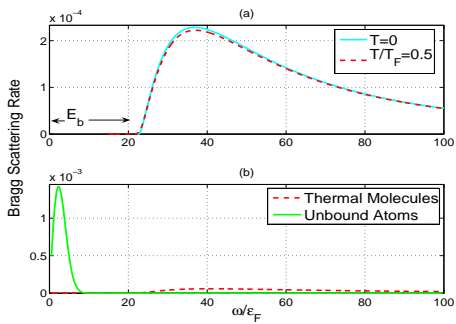}
\caption{Bragg scattering rate in units of
$\pi{\mathcal{I}}^2n/\epsilon_F$: from the molecular BEC, the thermal
molecules, and the free atoms.}
\label{MolBraggFig}
\end{figure}

    In Fig.\ \ref{MolBraggFig}, we plot the scattering rate as a function of
$\omega$ obtained from (\ref{MolBragg}) for a molecule energy
$E_b=-22.22\epsilon_F$ (corresponding to $k_Fa=0.3$) and for momentum transfer
$q=k_F$.  The scattering rate is shown in Fig.\ \ref{MolBraggFig} for a pure
molecular BEC at $T=0$ (a) and for thermal molecules at $T=0.5T_F$ above the
BEC critical temperature $T_c=0.218T_F$.  For $T=0$, the rate is given by
(\ref{BragBEC}) and the threshold for the scattering is given by
$\omega\ge-E_b+q^2/4m$ -- the minimum energy required to break a $p=0$
molecule and create two $\downarrow$ atoms with momenta ${\bf q}/2$ and $-{\bf
q}/2$.  The rate from the thermal molecules at $T=0.5T_F$ obtained from
(\ref{BragTherm}) does not differ significantly from that of the molecular BEC
at $T=0$.  This is because both temperatures are so low that there are
essentially no atoms present, as one sees in Fig.\ \ref{DensFig}.  The only
effect of finite temperature is then a small shift
$\omega\ge\min(-E_b+q^2/4m-qp/2m)\sim -E_b+q^2/4m-q\sqrt{k_BT/m}$ in the
threshold for scattering, from (\ref{BragTherm}).  Thus, the onset of
Bose-Einstein condensation of molecules has only minor effects on the Bragg
scattering rate since $|E_b|\gg k_BT_c$ in the BEC limit.

    In Fig.\ \ref{MolBraggFig} (b) we plot the scattering rate for
$T=2.5T_F$, at which most of the atoms are still bound as molecules but some
free atoms are present, as can be seen from Fig.\ \ref{DensFig}.  The Bragg
scattering consists of a signal down to $\omega=0$ from the free atoms, given
by (\ref{AtomBragg}), and a signal from the thermal molecules for
$\omega\gtrsim -E_b$, given by (\ref{BragTherm}).  Note that even though the
atomic signal at small $\omega $ is much larger than the molecular signal at
high $\omega$, the contribution to the f-sum rule from the molecules is much
larger than that from the free atoms.  For higher temperatures, the molecular
signal of course disappears with the vanishing molecule density, and the Bragg
scattering is given by that of the free atoms only.

\section{Strong coupling unitarity limit}

    The scattering length diverges, $k_F|a|\rightarrow \infty$, at the
Feshbach resonance.  In this unitarity limit, the interactions are strong and
considerably modify the properties of the gas.  An important characteristic of
the Fermi gas at unitarity is the presence of a pseudogap in the normal phase
close to $T_c$; i.e., the suppression of the single particle DOS around the
Fermi surface~\cite{Perali}.  A pseudogap has been observed experimentally for
high $T_c$ superconductors where it is presently subject to intense
investigations~\cite{Norman}.  Since a gap in the spectrum of single particle
excitations naturally leads to a gap in the low temperature Bragg spectrum
(Figs.\ \ref{wDepFig}-\ref{MolBraggFig}), we now examine whether the pseudogap
can be detected in Bragg spectroscopy.

    We calculate the Bragg spectrum at unitarity using a minimal many body
theory that includes the correct two-particle physics leading to the
resonance.  For the single channel problem, such a theory including pair
fluctuations in the ladder approximation was originally developed many years
ago~\cite{SRink} and later examined in great detail by several
authors~\cite{Randeria,Perali}.  This approach was then extended to the
multichannel case relevant for atoms interacting via a Feshbach resonance,
treating the bound state as a point
boson~\cite{holland,griffin,duine,bruunpethick}, and taking into account that
the bound state is a composite two-fermion object~\cite{BruunKolomeitsev}.

\begin{figure}
\includegraphics[width=0.8\columnwidth,
height=0.6\columnwidth,angle=0,clip=]{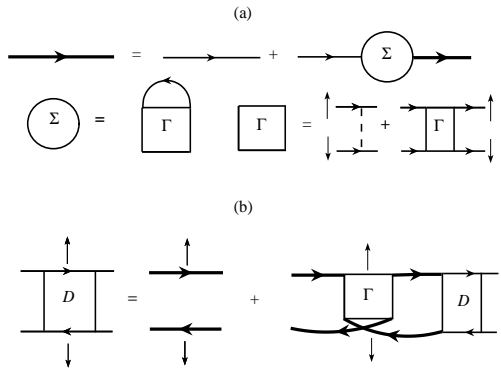}
\caption{(a) The atomic propagator in the ladder approximation for a broad
resonance where the Feshbach interaction is effectively instantaneous at
unitarity; (b) the spin-flip correlation function in a conserving ladder
approximation.}
\label{FeynFig}
\end{figure}

    The atoms are described by the atomic propagator
$G(q,z)^{-1}=G_0(q,z)^{-1}-\Sigma(q,z)$ with $G_0(q,z)^{-1}=z-\xi_q$.  The
atom self-energy is given in the ladder approximation by
\begin{equation}
 \Sigma(q,\omega_n)=\beta^{-1}{\rm Tr}[\Gamma({\mathbf{K}},{\mathbf{p}},
 {\mathbf{p}},\omega_n+\omega_m)G_0(k,\omega_m)]\label{Selfenergy}
\end{equation}
where ${\mathbf{K}}={\mathbf{q}}+{\mathbf{k}}$,
${\mathbf{p}}=({\mathbf{q}}-{\mathbf{k}})/2$, the trace denotes a sum over
Matsubara frequencies $\omega_m=i(2m+1)\pi T$ and integration over
${\mathbf{k}}$, and $\Gamma$ is the many-body scattering matrix depending on the 
center of mass ${\mathbf{K}}$ and relative momenta ${\mathbf{p}}$
of the two scattering particles~\cite{Fetter}.
In the limit of a broad resonance, the effective interaction mediated by the
Feshbach molecule can be regarded as instantaneous with a resonant scattering
length $|a|\rightarrow \infty$ at unitarity~\cite{bruunpethick}.  Neglecting
the frequency and momentum dependence of the effective interaction coming from
the molecular state, we recover the single channel theory scattering matrix
\begin{equation}
 \Gamma(K,z)^{-1}=\frac{m}{4\pi a}-\Pi(K,z).
 \label{Tmatrix}
\end{equation}
The pair propagator is, as usual,
\begin{equation}
 \Pi(K,z)=\int\frac{d^3k}{(2\pi)^3}\left(\frac{1-f-f'}{z-\xi-\xi'}
 +\frac{m}{q^2}\right).
 \label{PairProp}
\end{equation}
Here, $\xi=\xi_{{\mathbf{K}}/2+\mathbf{k}}$, $\xi'=\xi_{{\mathbf{K}}/2-\mathbf{k}}$,
$f=f(\xi)$, and $f'=f(\xi')$. The structure of
Eqs.~(\ref{Selfenergy})-(\ref{PairProp}) is indicated in Fig.\ \ref{FeynFig}
(a).

    To obtain the Bragg spectrum, we need to evaluate the atomic propagator at
frequency $\omega+i\eta$.  (From here on we do not write the infinitesimal
positive imaginary part of the frequency explicitly.)  From
(\ref{Selfenergy}), we obtain
\begin{gather}
 \Sigma(q,\omega)=\int\frac{d^3k}{(2\pi)^3}\int_{-\infty}^{\infty}
 \frac{d\epsilon}{2\pi}
 [f(\epsilon)A_0(k,\epsilon)\Gamma(k',\epsilon')-\nonumber\\
 n_B(\epsilon')A_{\rm m}(k',\epsilon')G_0(k,\epsilon)].
 \label{RSelfenergy}
\end{gather}
Here ${\mathbf{k}}'=\mathbf{{k+q}}$, $\epsilon'=\epsilon+\omega$, and
$A_0(k,\omega)=-2{\rm Im}G_0(k,\omega)$ and $A_{\rm m}(k,\omega)=-2{\rm
Im}\Gamma(k,\omega)$ are the spectral functions for non-interacting atoms and
the pair propagator, respectively.  We evaluate the real and imaginary parts of
$\Sigma$ directly from (\ref{RSelfenergy}).  We also calculate the real part
from the imaginary part using a Kramers-Kronig relation, and find that the two
calculations agree to a high accuracy, confirming the consistency of the
numerical calculations.

    In order to evaluate medium effects, we need to determine $T_c$ and the
chemical potential, $\mu$.  We determine $\mu$ by calculating the
thermodynamic potential $\Omega$ at unitarity in the ladder approximation.
The correction $\Delta\Omega$ to the ideal gas result can be expressed as an
integral over the phase of the on-shell scattering matrix
$\Gamma$~\cite{GMBUni}:
\begin{equation}
 \Delta\Omega=\int\frac{d^3Kd^3k}{(2\pi)^6}
 \frac{f(\xi)f(\xi')}
 {\rm{Im}\Pi(K,\xi+\xi')}\theta(K,\xi+\xi')
 \label{Omega}
\end{equation}
with $\xi$, $\xi'$ defined below (\ref{PairProp}).  The angle
$\theta(\omega)$ is defined by writing $\Gamma(\omega)=|\Gamma|e^{i\theta}$.
Equation (\ref{Omega}) reduces to the virial expansion for $\Omega$ at high
temperatures.  The chemical potential $\mu$ is then determined by
$-\partial\Omega/\partial_\mu=n$.

    In order to calculate the spin-flip correlation function consistently, one
needs in principle to choose a functional $\Phi[G]$, generate the self-energy
as $\Sigma = \delta \Phi/\delta G$, and then generate the correlation
functions from the kernel $\delta \Sigma/\delta G$~\cite{Baym}.  The diagrams
relevant for the ladder approximation are given in Fig.\ \ref{FeynFig} (b)
where all propagators are full Green's functions, $G$, including those in
$\Gamma$.  The conserving spin-flip correlation function thus obeys an
integral equation with $\Gamma$ and the full $G$'s.  The imaginary part of the
first term in this equation is
\begin{equation}
 {\rm Im}\mathcal{D}_0(q,\omega)=
 -\frac{1}{2}\int\frac{d^3k}{(2\pi)^3}
 \int_{-\infty}^{\infty}\frac{d\epsilon}{2\pi} AA'(f-f'),
 \label{ImD0Uni}
\end{equation}
with $A=A(k,\omega)$, $A'=A(k',\omegaø)$, and ${\mathbf{k}}'$ and $\xi'$
defined as in (\ref{RSelfenergy}).  The spectral function $A(k,\omega)=-2{\rm
Im}G(k,\omega)$ describes the excitations of the interacting atoms.

\subsection{Results}

    We calculate the Bragg scattering rate at the critical temperature $T_c$
where we expect the pseudogap effects to be most pronounced.  From the
thermodynamic potential (\ref{Omega}), we obtain $\mu(T)$ and can calculate the
critical temperature from the Thouless criterion $\Gamma^{-1}(0,0,T_c,\mu)=0$.
We then evaluate the atom self-energy from (\ref{RSelfenergy}) and from this
obtain the spectral function.

\begin{figure}
\includegraphics[width=0.9\columnwidth,
 height=0.7\columnwidth,angle=0,clip=]{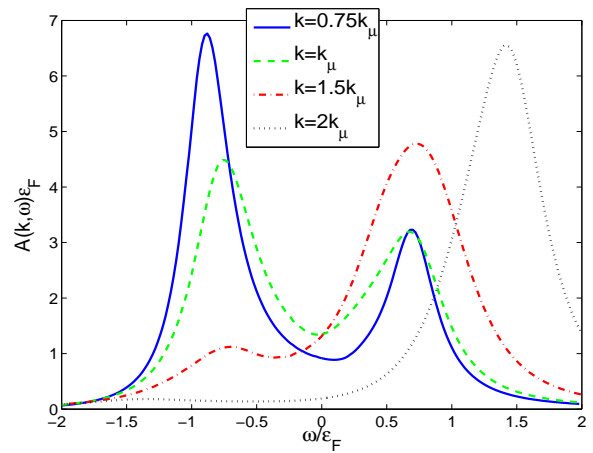}
\caption{The atom spectral function $A(k,\omega)$ at
unitarity and  $T=T_c$ for various momenta.}
\label{SpectralFig}
\end{figure}

    To understand the Bragg scattering rate at unitarity, we first analyze the
effects of the interactions on the single particle spectrum, described by the
spectral function, $A(k,\omega)$.  In Fig.\ \ref{SpectralFig}, we plot
$A(k,\omega)$ at $T_c$ for various momenta, measured in units of
$k_\mu=\sqrt{2m\mu}$.  We have chosen parameters corresponding to a resonant
interaction with $|k_{\rm F}a|=11.8\gg1$ and a negligible effective range.
For this set of parameters, we find $T_c\approx 0.26T_{\rm F}$ with
$\mu(T_c)\approx0.45\epsilon_F$ in good agreement with other BEC-BCS crossover
results based on a similar approximations~\cite{holland,Pieri}.  The spectral
functions obey the sum rule $\int A(\omega) d\omega/2\pi=1$ to a very good
approximation, providing another important check of the numerical calculations.

\begin{figure}
\includegraphics[width=0.9\columnwidth,
 height=0.7\columnwidth,angle=0,clip=]{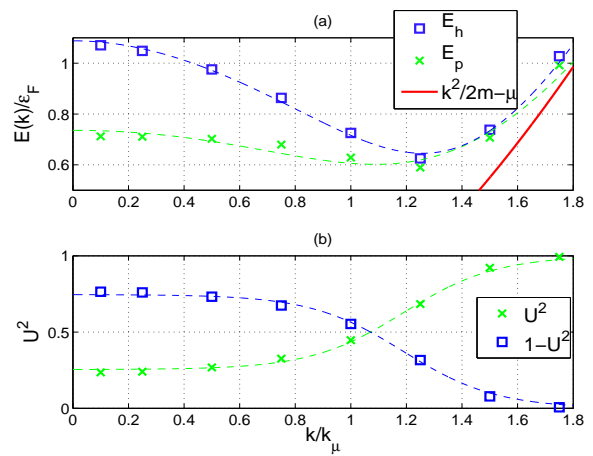}
\caption{(a) The positions $E_{\rm h}(k)$ and $E_{\rm p}(k)$ of the hole
and particle peaks of $A(k,\omega)$ as a function of $k$ at unitarity and for
$T=T_c$.  The solid line is the free particle energy $k^2/2m-\mu$.  (b) The
corresponding spectral weights of the particle and hole peaks.  The dashed
lines are fits to simple analytic functions.}
\label{PosFig}
\end{figure}

    In Fig.\ \ref{SpectralFig}, we clearly see the double peak structure of
$A(k,\omega)$ with the spectral weight at $\omega=0$ significantly suppressed
for all $k$.  This suppression is the characteristic feature of the
pseudogap.  We refer to the peak at $\omega<0$ as the {\em hole peak} at
frequency $-E_{\rm h}$, and the peak at $\omega>0$ as the {\em particle peak}
at $E_{\rm p}$.  The positions of the two peaks found by fitting to a double
Lorentzian (normalized such that its total spectral weight is $2\pi$) are
plotted in Fig.\ \ref{PosFig} (a).

    We see that the distance of the maxima to $\omega=0$ reaches a minimum:
${\rm min}E_{\rm h}(k)\sim {\rm min}E_{\rm p}(k)\sim0.6\epsilon_F$ at
$k\sim1.2k_\mu$, providing a qualitative value for the pseudogap.  The spectral
weight of the hole peak decreases and that of the particle peak increases with
increasing $k$.  This behavior is illustrated in Fig.\ \ref{PosFig} (b) where
we plot the spectral weights $U^2$ of the particle peak and the hole peak
$1-U^2$ obtained by fitting to the double Lorentzian.  For large $k/k_\mu$,
the spectral function approaches that of an ideal gas with a single peak at
$\omega=k^2/2m-\mu$ as expected.  The behavior of the spectral function at
unitarity as described here is in general agreement with the analysis of
Ref.~\cite{Perali}.  The spectral function behaves qualitatively as the BCS
spectral function
\begin{equation}
 A_{\rm BCS}(k,\omega)
 =v_k^22\pi\delta(\omega+E_k)+u_k^22\pi\delta(\omega-E_k),
\label{ABCS}
\end{equation}
with two peaks located around $\omega=0$.  There are important differences
from BCS theory though.  The two peaks for the spectral function in the
unitarity limit at $T_c$ have large widths (which are unequal for the particle
and hole peaks) and the spectral density is suppressed but finite at
$\omega=0$.  The pseudogap is a \emph{suppression} of spectral density and not
a real gap.  Also, the positions of the two peaks are not symmetric
around $\omega=0$ and the minimum gap is not at $k=k_\mu$ but slightly above,
as can be seen from Fig.\ \ref{PosFig}.

\begin{figure}
\includegraphics[width=0.9\columnwidth,
 height=0.7\columnwidth,angle=0,clip=]{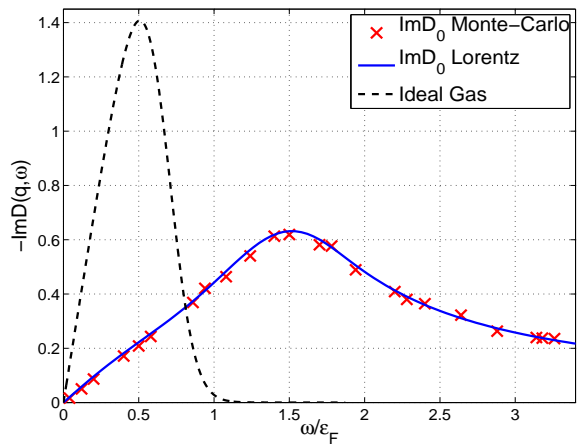}
\caption{$-{\rm{Im}}D_0(q,\omega)$ at unitarity and $T=T_c$ in units of
$n/2\epsilon_F$ as a function of $\omega$ for(color online).}
\label{BraggStrong}
\end{figure}

    We now ask whether the pseudogap behavior of the spectral function at
unitarity has consequences for the Bragg spectrum.  In Fig.\
\ref{BraggStrong}, we plot ${\rm Im}\mathcal{D}_0(q,\omega)$ for $q=k_\mu/2$
and $T=T_c$.  The $\times$'s are obtained from (\ref{ImD0Uni}) using the
spectral function calculated from (\ref{RSelfenergy}).  We evaluate the
(computationally demanding) multidimensional integrations using a Monte Carlo
routine.  The solid line in Fig.\ \ref{BraggStrong} is obtained by using the
double Lorentzian fit $A_L(k,\omega)$ for the spectral function in
(\ref{ImD0Uni}).  Here, the position of the hole $E_h(k)$ and particle
$E_p(k)$ peaks, the width and the weight $U^2(k)$ are calculated as functions
of momentum $k$ using approximate functions obtained from fitting
$A(k,\omega)$ to the double Lorentzian for various $k$.  The approximate
functions for $E_h(k)$, $E_p(k)$, and $U^2(k)$ are given as dashed lines in
Fig.\ \ref{PosFig}.  Once the approximate double Lorentzian form
$A_L(k,\omega)$ is obtained, the integration in (\ref{ImD0Uni}) is much faster
than when using the spectral function $A(k,\omega)$ obtained from the full
ladder calculation.  We see from Fig.\ \ref{BraggStrong} that the double
Lorentzian model $A_L(k,\omega)$ yields results for ${\rm
Im}\mathcal{D}_0(q,\omega)$ in good agreement with the full numerical
calculation based on $A(k,\omega)$.  From this, we conclude that the pseudogap
spectrum close to $T_c$ can be well described by a simple double peak spectral
function.

    Comparing to the ideal gas scattering rate for the same density and
temperature, also plotted in Fig.\ \ref{BraggStrong}, we see that interactions
suppress the spectral weight of ${\rm Im}\mathcal{D}_0(q,\omega)$ at low
$\omega$ and push it to higher frequencies.  This is a direct effect of the
pseudogap.  As opposed to the real gap in the $T=0$ Bragg spectrum, the
spectral density is however finite at small $\omega$.  This is as expected
since the single particle spectral density is suppressed but finite for
$\omega=0$ (Fig.\ \ref{SpectralFig}).  Also, $T_c\simeq0.26\epsilon_F$ is
comparable to the pseudogap $\sim0.6\epsilon_F$ so thermal excitations are
present.  From (\ref{ImD0Uni}), we see that that both effects make ${\rm
Im}\mathcal{D}_0(q,\omega)$ finite for small $\omega$.

    The approximation $\mathcal{D}_0(q,\omega)$ to the spin-flip correlation
function is not conserving and does therefore not obey the f-sum rule.  To
obtain a conserving approximation, we need to solve the integral equation in
Fig.\ \ref{FeynFig} (b).  This will change the quantitative form for
$\mathcal{D}(q,\omega)$.  The solution is however an iteration with the
$\Gamma$ matrix of $\mathcal{D}_0(q,\omega)$.  We therefore expect ${\rm
Im}\mathcal{D}(q,\omega)$ to have the same qualitative characteristics as
${\rm Im}\mathcal{D}_0(q,\omega)$, i.e., a suppressed but finite spectral
weight at low $\omega$.  Solving the fully self-consistent equation for the
spin-flip correlation function is beyond the scope of this paper
\cite{montecarlo}.

    In the BCS limit, we see that the emergence of a gap in the
single particle spectrum leads to a maximum in the Bragg scattering rate just
below $T_c$ (Fig.\ \ref{TDepFig}).  One could therefore expect a similar
maximum to appear at unitarity at a temperature $T^*$ defined as the
temperature where the pseudogap emerges.  However, the presence of a maximum
requires the spectral function to have well-defined sharp quasiparticle peaks
like in the BCS case (\ref{ABCS}).  The pseudogap spectral function at
unitarity (Fig.\ \ref{SpectralFig}) on the other hand has peaks of width
$\sim{\mathcal{O}}(\epsilon_F)$.  There will therefore be no maximum in the
Bragg scattering rate for $T\sim T^*$ due to the emergence of the pseudogap
\cite{slichter}.

    We conclude that the pseudogap present in the normal phase close to $T_c$
at unitarity leads to a significant suppression of the low frequency Bragg
scattering rate.  The rate remains however finite since the pseudogap is only
a suppression of the single particle spectral density, and because $T_c$ is
comparable to the value of the pseudogap.  The emergence of a pseudogap with
lowering $T$ does not give rise to a maximum in the scattering rate, as
opposed to the BCS case.  This is because the quasiparticles are strongly
damped at unitarity.

\section{Experiments}

    We briefly comment on the possible experimental observation of the effects
described in this paper.  A typical Fermi energy for trapped atomic gases is
$\sim{\mathcal{O}}(10^4)$Hz.  Thus, we require a frequency resolution of the
Bragg experiment $\ll 10^4$Hz.  Likewise, a typical Fermi momentum of the gas
is $k_F\sim{\mathcal{O}}(10^6)$m$^{-1}$ which is the same order of magnitude
as the wave numbers for optical lasers.  This means that the resulting angle
between two laser beams needed to obtain a momentum transfer of $\sim k_F$ is
reasonable.  The results presented in this paper are for a homogenous system.
From the Thomas-Fermi approximation, we expect the presence of the trapping
potential to smooth out features like the gap in the Bragg spectrum.  To avoid
such smoothening, one could focus the laser beams on a smaller section of the
cloud where the gas can be regarded as homogeneous.  In this case, the results
presented in the present paper become directly applicable.

%-Check with smaller Gamma. Maybe sum-rule better?

\section{Acknowledgments}

    We are grateful to Cindy Regal and Chris Pethick for helpful discussions.
We thank the Aspen Center for Physics where parts of this research were
carried out.  This research was also supported in part by NSF Grants
PHY03-55014 and PHY05-00914.

\end{document}